\documentstyle[12pt,aps]{revtex}
\begin{document}

\title{The halo of the exotic nucleus $^{11}$Li: a single Cooper pair}
\author{F. Barranco$^1$, R. A. Broglia$^{2,3,4}$, G. Col\`o$^{2,3}$, 
E. Vigezzi$^{2,3}$ \\
$^1$Escuela de Ingenieros Industriales, \\Universidad de Sevilla, Camino de los
Descubrimientos, 41092 Seville, Spain\\
$^2$Department of Physics, University of Milan, Via Celoria 16, 20133, Milan,
Italy\\
$^3$INFN, Sezione di Milano, Milan, Italy\\
$^4$The Niels Bohr Institute, University of Copenhagen, \\Blegdamsvej 17, 2100
Copenhagen, Denmark}

\maketitle

\bigskip
\bigskip
\begin{abstract}
{If neutrons are progressively added to a normal nucleus, the Pauli principle 
forces them into states of higher momentum. When the core becomes
 neutron-saturated, the nucleus expels most of the wavefunction of the last
 neutrons outside to form a halo, which because of its large size can have 
lower momentum. It is an open question how nature stabilizes such a fragile 
system and provides the glue needed to bind the halo neutrons to the core. 
Here we show that this problem is similar to that of the instability 
of the normal 
state of an electron system at zero temperature solved by Cooper, solution
 which 
is at the basis of BCS theory of superconductivity. By mimicking this 
approach using, aside from the bare nucleon-nucleon interaction, 
the long wavelength vibrations of the nucleus $^{11}$Li, 
the paradigm
of halo nuclei, as tailored glues of the least bound neutrons, we are able
to obtain a unified and quantitative picture of the observed properties 
of $^{11}$Li.}
\end{abstract}
\newpage

Aside from ``dark matter''~\cite{Pea99}, atomic nuclei, little 
droplets
made out of protons and neutrons 10 femtometer across~\cite{Boh6975}, make up
a sizable fraction of the present mass of the universe. Research into the 
structure of atomic nuclei concentrates largely on the limits of the 
nuclear stability~\cite{Oga99}, where new physics is
expected to be, in particular at the limits of neutron and 
proton number defining the so called drip lines in the chart of nuclides. 
The most exotic nuclei~\cite{VARENNA}, first produced in the laboratory only few years ago, are 
those that lie just within the drip lines, on the edges of nuclear stability. 
Of these, the atomic nucleus $^{11}_3$Li$_8$, containing 3 protons and 8 
neutrons, is rightly the most famous and better studied and provides the 
cleanest 
example of halo nuclei to date~\cite{HANSEN,Kob88,Tan96,Ann90,Arn92,Esb92,Bar96}.

In halo nuclei, some of the constituent neutrons or protons venture beyond the
drop's surface and form a misty cloud or halo. Not surprisingly, these
extended nuclei behave very differently from ordinary (``normal'') nuclei
lying along the so-called stability valley in the chart of nuclides. In 
particular, they are larger than normal nuclei of the same mass number, and 
they interact with them with larger cross sections as well. 
In the case of $^{11}$Li, 
the last two neutrons are very weakly bound. 
Consequently, these neutrons need very little energy to move away
from the nucleus. There they can remain in their ``stratospheric'' orbits, 
spreading out and forming a tenuous halo. If one neutron is taken away from
$^{11}$Li, a second neutron will come out immediately, leaving behind the
core of the system, the ordinary nucleus
$^{9}$Li. This result testifies to the fact that pairing, the attraction 
correlating pairs between the least bound particles in a system, plays
 a central role in the stability of $^{11}$Li. 

It is well known that pairing can radically affect the properties of a 
many-body system. 
In metals, 
pairing between the electrons gives rise to 
superconductivity~\cite{SCHRIEFFER}. 
In a dilute neutron gas, pairing
can influence the properties of neutron stars, cold remnants of fierce
supernova explosions~\cite{structure}. Acting on a liquid made out of
atoms of the lighter isotope of helium ($^3_2$He$_1$) it leads to
superfluidity~\cite{Leg89}.
In nuclei,
it controls almost every aspect
of nuclear structure close to the ground state 
and determines, to a large extent,  which nuclei are stable and which are
not~\cite{Boh6975,Wal99}.

The basic experimental facts which characterize $^{11}$Li and which are also 
of  particular relevance in connection with pairing in this system are: a) 
$^9 _3$Li$_6$ and $^{11}_3$Li$_8$ are stable, $^{10}_3$Li$_7$ is not, b) the 
two-neutron separation energy in $^{11}$Li is only $S_{2n}$=0.294$\pm$0.03 
MeV~\cite{Tan96} as compared with values of 10 to 30 MeV in stable nuclei,
 c) $^{10}$Li displays s- and p-wave resonances at low energy, their centroids 
lying within the energy range 0.1-0.25 MeV and 0.5-0.6 MeV 
respectively~\cite{Zin95},
 d) the mean square radius of $^{11}$Li, $\langle r^2 \rangle^{1/2}$=3.55$\pm$
0.10 fm~\cite{Koba89,Toste96,Han_boh}, is very large as compared to the value 2.32$\pm$0.02 fm 
of the $^9$Li core, and testifies to the fact that the neutron halo must have 
a large radius ($\approx$6-7 fm),
 e) the momentum distribution of the halo neutrons is found to be exceedingly 
narrow, its FWHM being equal to $\sigma_{\bot}=48\pm 10$ MeV/c for the
 (perpendicular) distribution observed in the case of the break up of 
$^{11}$Li on $^{12}$C, a value which is of the order of one fifth of that 
measured 
during the break up of normal nuclei~\cite{Kob88,Tan96},
 f) the ground state of $^{11}$Li is a mixture of configurations where the two
 halo nucleons move around the $^9$Li core in $s^2-$ and $p^2-$configurations 
with almost equal weight~\cite{aoi97,simo99}.

Before discussing the sources of pairing correlations in $^{11}$Li, we shall
study the single-particle resonant spectrum of $^{10}$Li. The basis of
(bare) single-particle states used was determined by calculating the
eigenvalues  and eigenfunctions of a nucleon moving in a Saxon-Woods
potential with spin-orbit and symmetry term  (cf. \cite{Boh6975}, Vol. I,
Eqs. (2.281) and (2.282)). The continuum states of this  potential  were
calculated by solving the problem in a box of radius equal to 40 fm, 
chosen so as to make the results associated with $^{10}$Li and $^{11}$Li
discussed below, stable.  While mean field theory predicts the orbital
$p_{1/2}$ to be lower than the $s_{1/2}$ orbital (cf. Fig. 1, I(a)),
experimentally the situation is reversed. Similar parity inversions have
been observed in other isotones of $_3^{10}$Li$_7$, like e.g.
$^{11}_4$Be$_7$. Shell model calculations testify to the fact that the 
effect of core excitation, in particular of quadrupole type, play a
central role in this inversion \cite{Sag93},(cf. also \cite{Vin95}). 
In keeping with this result,
we have studied the effect the coupling of the $p_{1/2}$ and $s_{1/2}$
orbitals of $^{10}$Li to monopole-, dipole- and quadrupole-vibrations of 
the $^9$Li core has on the properties of the 1/2$^+$ and 1/2$^-$ states
of this system.
In this study we have used vibrational states calculated by diagonalizing,
in the random phase approximation (RPA), a multipole-multipole separable
interaction taking into account the contributions arising from the excitation
of particles into the continuum states.
The coupling strengths were adjusted so as to reproduce the position and
transition probabilities of vibrational states in the normal nucleus $^{10}$Be
\cite{Ajzenberg}.
Unperturbed particle-hole
excitations up to 70 MeV have been included in the calculations 
and phonon states
up to 50 MeV have been considered. Within this space there are of the order
of $10^2$ dipole states and about the same amount of 
quadrupole states, exhausting
the associated energy-weighted sum rules.
The resulting vibrational states were coupled to 
the particle states making use of the associated transition densities
(formfactors) and 
particle-vibration coupling strengths.
A Skyrme-type effective interaction (SLy4) was used
to calculate the 
monopole linear response. The corresponding solutions were obtained 
in coordinate space making use of a mesh extending up to a radius of 80 fm.
The monopole response exhausts 94\% of the EWSR 
considering the summed contributions up to 40 MeV of excitation energy.
This response function was discretized in bins of 300 keV and the resulting
states coupled to the single-particle states making use of the corresponding
transition densities and particle-vibration coupling strengths.
Similar calculations were performed to determine the properties of
the $L$=0,1 and 2-vibrational states of $^{11}$Li needed 
in connection with the discussion 
of the pairing induced interaction carried out below.
In this case  the strength of the separable dipole 
interactions has been slightly changed from the value used in the previous 
calculation so as to provide 
an overall account of the experimental dipole response in $^{11}$Li
\cite{Sackett,Zin97}.
For the quadrupole strength the same value used 
before was employed. Because the calculations have been carried out 
on the physical (correlated ) $^{11}$Li ground state, the transitions 
associated with the vibrational states involving the $p_{1/2}$ and the $s_{1/2}$
states have been renormalized with the corresponding occupation numbers 
resulting from the full diagonalization. 

In the study of
the single-particle resonances of $^{10}$Li we have considered not only 
the particle-vibration coupling
vertices associated with effective-mass-like diagrams (upper-part graph 
of Fig. 1, I(b)) and leading to attractive (negative) contributions to the
single-particle energies, but also those couplings leading to Pauli
principle (repulsive) correction processes associated with diagrams
containing two-particles, one-hole and a vibration in the intermediate
states (lower-part diagram of Fig. 1, I(b)). Because of such Pauli correction
processes, the $p_{1/2}$ state experiences an upward shift in energy,
arising from the coupling of this orbital to the $p_{3/2}$ hole-state through
quadrupole vibrational states, in keeping with the fact that the ($p_{1/2}
p_{3/2}^{-1}$) particle-hole excitation constitutes an important component of
the quadrupole vibration wavefunction. As a consequence, the $p_{1/2}$
state becomes unbound, turning into a low-lying resonance with centroid 
$E_{res} \approx $ 0.5 MeV.
Due to the coupling to the vibrations the $s-$states are instead shifted
downwards. In fact, in this case there are essentially no (repulsive) 
contributions arising from the Pauli correction processes. On the other hand,
(attractive) effective-mass-like processes with intermediate states consisting
of one particle
plus a vibrational state of the type ($s_{1/2} \times 0^+$), 
($p_{1/2} \times
1^-$) and ($d_{5/2} \times 2^+$) lead to a virtual state with  $E_{virt} =
0.2 $ MeV (cf. Fig. 1, I(b)), the coupling  to quadrupole vibrations
providing by far the largest contribution to the total energy shift.
The important difference between the distribution of the  single-particle
strength associated with the resonant state $p_{1/2}$ and the virtual
state $s_{1/2}$ can be  observed in Fig. 1, I(c), where the partial cross section
$\sigma_l$ for neutron elastic scattering off $^9$Li is shown. While
$\sigma_p$ displays  a clear peak at 0.5 MeV, $\sigma_s$ is a smoothly
decreasing function of the energy. On the other hand, a small increase in the
depth of the potential felt by the $s-$neutron will lead to a (slightly)
bound state, hence the name of virtual \cite{Landau}.

At the basis of the variety of pairing phenomena observed in systems so
apparently different as atomic nuclei and metals 
is the formation of Cooper pairs~\cite{boh56,bardeen}. In the case of metals,
Cooper pairs arise from the combined
effect of Pauli principle and of the exchange of lattice vibrations (phonons) 
between pairs of electrons (fermions) moving in time reversal states lying
close to the Fermi energy. In the superconducting or BCS state
of metals~\cite{SCHRIEFFER}, Cooper pairs are strongly overlapping. In fact, there are on 
average 10$^6$ pairs which have their centers of mass falling within the 
extent of a given pair wavefunction. In spite
of the modest number of Cooper pairs present in the ground state of atomic 
nuclei ($\le$ 10), BCS theory gives a quite accurate description
of pairing in nuclei~\cite{Boh6975,Belyaev}, the analogy between the pairing
gap typical of metallic superconductors and of atomic nuclei being very much to
the point \cite{Pines}.
On the other hand, the finiteness  of the nucleus introduces in the BCS 
treatment 
of pairing important modifications (quantal size effects (QSE),
\cite{Belyaev},\cite{Kubo}-\cite{Anderson}). 
In particular, while in the infinite system the existence of a bound state
of the (Cooper) pair happens for an arbitrarily weak interaction 
\cite{boh56},
 in the nuclear case this phenomenon takes place only if the strength
of the nucleon-nucleon potential is larger than a critical value connected 
with the discreteness of the nuclear spectrum. In fact, calculations carried out
making use of a particularly successful parametrization of the (bare)
potential (Argonne potential \cite{Argonne}), 
testify  to the fact that the nuclear forces
are able to bind Cooper pairs in open shell nuclei (leading to sizable
pairing gaps (1-2 MeV) \cite{Barranco1}), 
but not in closed shell nuclei, the most
important contributions to the nucleon-nucleon (pairing) interaction
arising from high multipole components of the force \cite{Belyaev}. 

The situation
is however quite different for the "open  shell" nucleus $^{11}$Li. In fact,
allowing the two neutrons to interact 
through the Argonne potential 
produces almost no  mixing between $s-$ and $p-$waves, 
but essentially it only  shifts the energy of 
the unperturbed (resonant) states $s^2_{1/2}(0)$ and 
$p^2_{1/2}(0)$ by about 80 keV without giving rise to a bound
system.
Similarly, making use of the matrix elements of 
the same nucleon-nucleon potential in connection with the BCS equations does not
lead to a solution but the trivial one of zero pairing gap  ($\Delta_{\nu} =0,
U_{\nu}V_{\nu}$ = 0) \cite{Belyaev}.
At the basis of this negative result is the fact that the most important
single-particle states allowed to the halo neutrons of $^{11}$Li to correlate
are the $s_{1/2}$, $p_{1/2}$ and $d_{5/2}$ orbitals. Consequently the two 
neutrons are not able, in this  low-angular momentum phase space, to profit
fully from the strong force-pairing interaction, as only 
the components of multipolarity $L=0 $, 1 and 2 of this force
are effective in $^{11}$Li
because of angular momentum and parity conservation \cite{reason}.

In keeping with this result and with those of  ref. \cite{Barranco}, 
and making use of the fact that
$^{11}$Li displays low-lying collective vibrations \cite{Sackett,Zin97},
one can posit that the exchange of these vibrations
between the two neutrons is the main source of pairing available 
to them to correlate. In fact, 
allowing the two neutrons to both exchange phonons (induced
interaction, Fig. 1, II(a)), as well as to emit and later reabsorb
them ( self-energy correction, Fig. 1, I(b)), leads to a bound
(Cooper) pair, the lowest eigenstate of the associated 
secular matrix being  $E_{gs}=$
-0.270 MeV. Adding to the induced interaction the nucleon-nucleon Argonne
potential  one obtains $E_{gs}$ = -0.330 MeV, and thus a two-neutron
separation energy quite close to the experimental value.
 Measured from the unperturbed energy
 of a pair of neutrons in the lowest state calculated for $^{10}$Li, namely
the $s$-resonance 
($E_{unp}=2E_{s_{1/2}}=$ 400 keV, cf. Fig. 1,I(b)), 
it leads to a pairing correlation
 energy $E_o=E_{unp}-E_{g.s.}=$ 0.730 MeV (cf. Fig. 1, II(b)).
From the associated two-particle ground state wavefunction 
$\Psi_0(\vec r_1, \vec r_2)(\equiv \langle \vec r_1, \vec r_2 |0^+ \rangle)$, 
one obtains a momentum distribution 
(whose FWHM is $\sigma_{\bot}=56$ MeV/c, for $^{11}$Li on $^{12}$C)
 and ground state occupation 
probabilities of the two-particle states $s_{1/2}^2$(0), $p_{1/2}^2$(0) and
$d_{5/2}^2(0)$
 (0.40, 0.58 and 0.02 
respectively, cf. Fig. 1,II(b))
which provide an overall account of the experimental findings.
The radius of the associated single-particle distribution is 7.1 fm.
 Adding to this density
 that of the core nucleons one obtains the total density 
of $^{11}$Li. The associated mean square radius (3.9 fm) is
slightly larger than the experimental value.

The spatial structure of the Cooper pair described by the wavefunction 
$\Psi_0(\vec r_1, \vec r_2)$ is displayed in Fig. 2. 
The mean square radius of the center of mass of the two neutrons is
$\langle r_{cm}^2 \rangle^{1/2}$ = 5.4 fm. 
This result testifies to the importance the correlations have
in collecting the small (enhanced) amplitudes of the uncorrelated 
two-particle configuration $s_{1/2}^2(0)$ in the region between 4 to 5
fm, region in which the $p_{1/2}^2(0)$, helped by the centrifugal barrier,
displays a somewhat larger concentration (Fig. 3).
From the above results, it emerges that the exchange of vibrations 
between the least bound neutrons leads to a (density-dependent) pairing 
interaction acting essentially only outside the core (cf. also ref. \cite{Esb91}).
To be noted that the long wavelength behaviour of these vibrations
is connected with  the excitation of the neutron halo,  the large size of which
not only makes the system easily polarizable but provides also the
elastic medium through which the loosely bound neutrons exchange vibrations with
each other \cite{Pauli}.

The average mean square distance between the halo neutrons is
$<r_{12}^2 >^{1/2} \approx $ 9.2 fm, in keeping with the fact that the 
coherence
length \cite{SCHRIEFFER} associated with Cooper pairs in nuclei is larger
than the nuclear dimensions  thus preventing the possibility of a nuclear
supercurrent (cf. e.g. \cite{Boh6975} Vol. II, p. 398). 
On the other hand, this value of
$<r_{12}^2 >$  does not prevent the two correlated neutrons to be close together.
The corresponding (small) probability (cf. Fig. 2) being much larger than
that associated with the uncorrelated neutrons (cf. Fig. 3).

Similar results as those reported above are obtained solving the BCS equation 
for the two-neutron system making use of the matrix elements used
in the diagonalization, sum of 
those of the nucleon-nucleon Argonne potential and those of the
induced interaction.
In this case, the correlation energy is $E_o$= 0.7 MeV, 
the separation energy of the two neutrons becoming $S_{2n}=$ 0.3 MeV. The 
radial structure of the projected BCS wavefunctions 
$\sum _{\nu >0}(V_{\nu}/U_{\nu}) \varphi_{\nu} (\vec r_1) \varphi_{\nu} 
(\vec r_2)$ displays a spatial structure quite similar to 
$\Psi_0(\vec r_1, \vec r_2)$, the admixture of s- , p- and d- two particle 
configurations being now 46$\%$ and 51$\% $ and $3\%$
respectively. The coherence 
length $\xi$, that is the mean square distance between the two neutrons 
forming the Cooper pair, is in this case, 
$\langle r_{12}^2\rangle ^{1/2}$ = 7.8 fm.

Arguably, the understanding of halo nuclei is the single, most important issue
 of nuclear structure research still awaiting a satisfactory explanation. 
We have shown that a substantial advance towards this goal is made by properly
 characterizing the role the surface of the system plays in renormalizing 
the bare nucleon-nucleon potential and the single-particle motion
of the nucleons. 
In fact, we find compelling evidence which testifies 
to the fact that the mechanism which is at the basis of the presence of a low 
density halo in $^{11}$Li, is the coupling of 
weakly bound nucleons to long wavelength vibrations of the system leading to 
pairing instability and thus to a bound system. 
This result suggests a general strategy for designing nuclei with a large
excess of neutrons or protons which may prove valuable in the current
exploration of the drip line of the chart of nuclides, and thus in the design
of new nucleon species: select those systems which display, under an increase
of the excess of one type of nucleons, a marked softening of the long
wavelength linear response. It is likely that such systems could venture
far inside the region delimited by the drip lines, in keeping with 
the fact that the associated vibrational modes are expected to be an important
source of (induced) pairing interaction, and thus to contribute significantly
to the stability of the system.

We wish to thank G. Gori for technical help.

\bigskip
\bigskip

Correspondence and request for materials should be addressed to RAB (e-mail:
broglia@mi.infn.it)

\newpage

{\bf Caption to the figures}

\vskip 4mm

{\bf Fig. 1}

(I) {\bf Single-particle neutron resonances in $^{10}$Li}.
 In (a) the position of the levels $s_{1/2}$ and $p_{1/2}$ calculated
 making use of mean field theory is shown (hatched area and 
thin horizontal line respectively). 
The coupling of a single-neutron
(upward pointing arrowed line) to a vibration (wavy line) calculated 
making use of the Feynman diagrams displayed in (b) 
(schematically depicted 
also 
in terms  of either solid dots (neutron) or open circles (neutron hole) 
moving in a single-particle level around or in 
the $^9$Li core (hatched area)), leads to conspicuous shifts in the energy 
centroid of the 
$s_{1/2}$ and $p_{1/2}$ resonances (shown by thick horizontal lines) and 
eventually to an inversion in their sequence.
In (c) we show the calculated partial cross section 
$\sigma_l$ for neutron elastic scattering off $^9$Li.  

(II) {\bf The two-neutron system $^{11}$Li}. 
We show in (a) the mean-field picture 
of $^{11}$Li, where two neutrons (solid dots) move in time-reversal states 
 around the core $^9$Li (hatched area) in the $s_{1/2}$ resonance leading to 
an unbound $s^2_{1/2}(0)$ state where the two neutrons are coupled to 
zero angular momentum. The exchange of vibrations 
between the two neutrons shown in the upper part of the figure leads to a 
density dependent interaction which, added to the 
nucleon-nucleon interaction, correlates the two-neutron system leading
to a bound state $|0^+ \rangle$, where the two 
neutrons move with probability 
0.40, 0.58 and 0.02
in the two-particle configurations $s^2_{1/2}(0)$,  $p^2_{1/2}(0)$ and
$d^2_{5/2}(0)$ respectively.
 
\vskip 4mm

{\bf Fig. 2}

{\bf Spatial structure of two-neutron Cooper pair}.
The modulus squared wavefunction
$|\Psi_0 (\vec r_1,\vec r_2)|^2=|\langle \vec r_1,\vec r_2|0^+ \rangle|^2$
 (cf. Fig. 1, II (b)) describing the motion of the two halo 
neutrons around the 
$^9$Li core (normalized to unity and multiplied by 16$\pi^2r_1^2r_2^2)$ is 
displayed as a function of the cartesian coordinates 
$x_2 = r_2 \cos(\theta_{12})$
and $y_2=r_2 \sin(\theta_{12})$ of particle 2, for fixed value of 
the position of particle 1 
($r_1$=2.5, 5, 7.5 fm) represented in the right panels by a solid dot, while
the core $^9$Li is shown as a red circle. The numbers appearing on the $z$-axis
of the three-dimensional plots displayed on the left side of the figure are 
in units of fm$^{-2}$.

\vskip 5mm

{\bf Fig. 3 }
{\bf Spatial distribution of the pure two-particle configurations 
 $ \mbox{\boldmath $s_{1/2}^2$}$ (0) 
and $ \mbox{\boldmath $p_{1/2}^2$}$(0)} 
as a function of the $x$- and  $y$-coordinates of particle
 2, for a fixed 
value of the coordinate  of particle 1 ($r_1$=5 fm). For more details cf.
 caption to Fig. 2.


\begin{thebibliography}{99}
\bibitem{Argonne} This nucleon-nucleon potential has been designed 
\cite{Wiringa} so as to accurately reproduce the nucleon-nucleon phase shifts.
\bibitem{reason} This is the reason why the Cooper pair problem and the BCS 
treatment of pairing in the case of $^{11}$Li looks (qualitatively)
different from pairing in nuclei lying along the stability valley, and 
more similar to pairing in metals \cite{SCHRIEFFER},\cite{boh56}, 
or even better
in doped fullerides \cite{Lieber}-\cite{Breda}.
\bibitem{Pauli} In keeping with the fact that the vibrational states of $^{11}$Li
are built out of
excitations which occupy, to some extent, the same
particle states occupied by the loosely bound neutrons we are studying,
the corresponding particle-vibration matrix elements have been corrected
from Pauli violating contributions.
In particular, the reduction factors aassociated with
the  matrix elements $<s_{1/2}\times 1^-|H_c|p_{1/2}>$,
$<s_{1/2}\times 0^+|H_c|s_{1/2}>$ and
$<p_{1/2}\times 0^+|H_c|p_{1/2}>$ of the particle-vibration
coupling Hamiltonian $H_c$ the reduction factor are
0.68, 0.25 and 0.25 respectively.
\end{thebibliography}
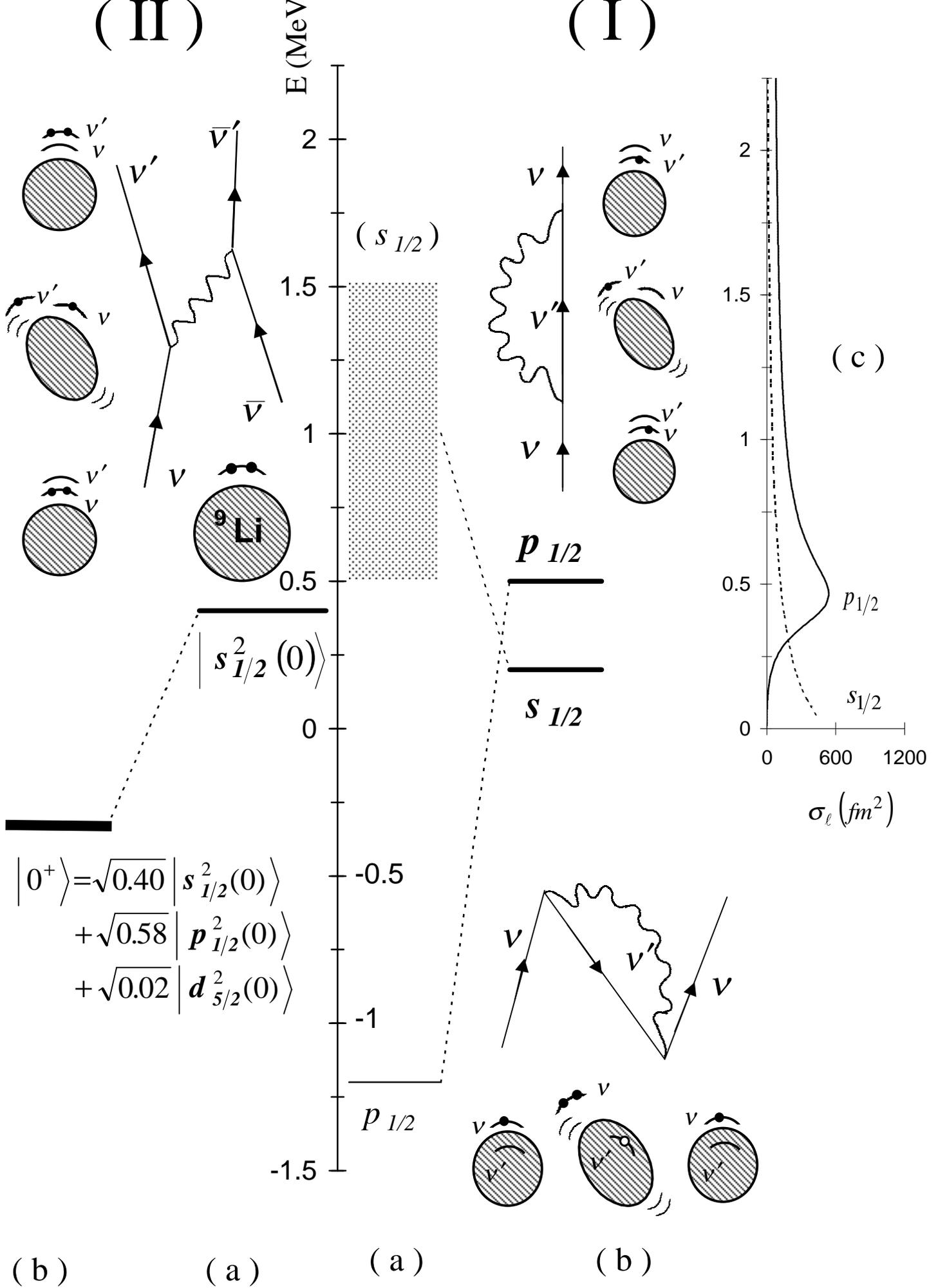
\end{document}